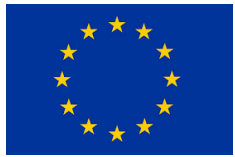

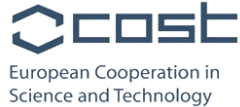

# EU COST ACTION ON FUTURE GENERATION OPTICAL WIRELESS COMMUNICATION TECHNOLOGIES - NEWFOCUS CA19111

*A White Paper*

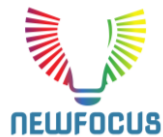

https://www.newfocus-cost.eu/action/

**July 2022**

# Contents

# 1. Introduction

Horizon Europe the EU's ninth multiannual Framework Programme for research and innovation is the largest programme of its kind anywhere in the world with a total budget of €95 billion [1]. It promises more breakthroughs, discoveries, and world-firsts by taking great ideas from the lab to the market. The new framework will focus on four pillars:

I. **Pillar 1** – Excellent Science, aims to increase the EU's global scientific competitiveness.
II. **Pillar 2** - Global Challenges and European Industrial Competitiveness, which covers (i) *Health; (ii) culture, creativity, and inclusive society; (iii) civil security for society; (iv) digital, industry and space; (v) climate, energy, and mobility; (vi); and (vii) food, bioeconomy, natural resources, agriculture, and environment*
III. **Pillar 3** - Innovative and Inclusivity in Europe, which aims to make Europe a frontrunner in market-creating innovation via the European Innovation Council.
IV. **Horizontal Pillar** - Widening participation and strengthening the European Research Area.

A significant part of Pillar II will be implemented through institutionalized partnerships, particularly in the areas of Mobility, Energy, Digital, and Bio-based economy, which will also have separate work programmes.

The EU COST Action **NEWFOCUS** [2] is focused on investigating radical solutions with the potential to impact the design of future wireless networks. It aims to address some of the challenges in OWC and establish it as an efficient technology that can satisfy the demanding requirements of backhaul and access network levels in 5G networks. This also includes the use of hybrid links that associate OWC with radiofrequency or wired/fiber-based technologies. The focus of this **White Paper** is on the use of optical wireless communication (OWC) as enabling technology in a range of areas outlined in HE's Pillar II including Health, Manufacturing, Intelligent Transportation Systems (ITS), Unmanned Aerial Vehicles and Network and Protocol.

The fifth and sixth generation (5/6G) wireless networks and beyond are introducing radical changes to ensure the full realization of the smart IoT and Internet of everything paradigms capabilities people, machine, devices, etc. This will require radically different communication networks, which are dynamic, sustainable, intelligent, simple, reliable, and energy efficient. The first paper on "*Optical Wireless Communications – The Context*" by Z Ghassemlooy, A-M Khalighi, S Zvanovec, N Stevens, Luis Nero Alves, and Amita Shrestha discusses this and focuses on the complementary optical wireless communication technology.

ITS combines data communication, positioning, computing, and automation technologies to improve traffic management, safety on the roads, and transportation efficiency, where wireless technologies play an important role. The next two papers on "*Visible Light Communications for Intelligent Transportation Systems*" by M Uysal and "*On the Use of Visible Light Communications in Intelligent Transport Systems: From Road Vehicles To Trains*" by A M Vegni gives an overview of ITS, focusing on the concept and advantages of visible light communications and optical camera communications in ITS including vehicular and train communications, the start of the art in research and development, as well as challenges and future directions.

As part of 5G and beyond wireless networks, unmanned aerial vehicles (UAVs), which are becoming more mature in technology and with reduced costs, are considered as mobile communication terminals with a high capacity, whereas radio frequency-based wireless communications may not be able to support it. The paper on "*Unmanned Aerial Vehicles with Lightwave Technology for 6G Networks Authors*" by P D Diamantoulakis, et al, discusses the concept of energy-autonomous UAV-based communications with simultaneous light wave information and power transfer, the state of the art, challenges, and future research works including quantum key distribution via UAVs.

There are many challenges facing the manufacturing industry, where reconfiguring the facilities is difficult and expensive when considering wired-based network connectivity. The use of smart technologies with wireless communications systems is helping manufacturers to re-arrange the layout of the factory floor in a flexible way to meet changing requirements and products, thus improving productivity. The paper on "*Optical Wireless Communications Technologies for Smart Manufacturing in Industry 4.0*" by B Ortega and V Almenar outlines the 5 and 6 G technologies in industry and the role of OWC in providing data communications and highly precise indoor localization.

Virtual reality is radically changing the interactions between humans and the outside world by building a synthetic virtual environment to mimic the real world, which can be used for social sharing, video streaming, and 6 degree-of-freedom content streaming, etc. The main issues in virtual reality are wired devices, limited mobility, insufficient bandwidth, and real-time support. The paper on "*Virtual Reality*" by O Bouchet introduces the state of the art in virtual reality and issues, followed by the role of OWC in providing a very high-speed wireless technology. The last paper on "*IPv6-based IoT in 2025*" by L Ladid discusses how IPv6 can enable and sustain the growth rate of the IoT and offers a future-proof solution.

Finally, we hope that this first White Paper will serve as a valuable resource for a better understanding of the OWC technology highlighting some of the features, issues, and research works carried out in emerging application areas as well as promoting further research, and development in the successful deployment of OWC systems as a protuberant complementary to RF-based technologies in the 5&6G and beyond heterogeneous wireless networks.

If you would like to participate in NEWFOCUS activities please contact the Steering and Management Committees members as well as Working Groups Chairs, for details see [2].

[1] https://www.horizon-eu.eu/
[2] https://www.newfocus-cost.eu/action/

M-A Khalighi, *Chair of Newfocus*
Z Ghassemlooy, *Vice-Chair of Newfocus*

# 2. OPTICAL WIRELESS COMMUNICATION – THE CONTEXT

Z Ghassemlooy, M-A Khalighi, S Zvanovec, N Stevens, L N Alves, and A Shrestha

I. INTRODUCTION

Globally, the society that we are living in is changing rapidly and becoming more data-centric, data-dependent, and fully automated with the aim of improving productivity and the quality of life. In a fully automated and intelligent world the Internet of Things (IoT) will play a critical role in connecting billions of physical devices in the fabric of society (i.e., people, machines, homes, offices, industry, cities, environments, etc) to sense, collect and exchange data, and improving interactions between devices across various sectors, such as manufacturing, the connected home, transportation, medical, and agriculture. The widespread uptake of Industry 4.0 and the emerging smart environments (cities, factories, offices, homes, etc.) requires wireless IoT devices that can collect data and transmit them to a central location via telecommunication networks (particularly wireless networks). Within this context, fifth and sixth generation (5/6G) wireless networks are aiming to offer full realization of the IoT paradigm with machine learning capabilities for connecting not just people, but also people-to-vehicles, people-to-devices, machine-to-machine, sensors, wearables, cloud resources, robots, etc. Therefore, in such challenging environments, there is the need for radically new telecommunication networks with key features of utilizing new spectrums, disruptive technologies, machine learning, enabling technologies, energy efficiency, sustainability, etc. Enabling technologies and solutions include millimetre wave (mmW) (mostly unregulated bands up to 90 GHz) and terahertz (THz) bands; cell-free networks; cognitive radio systems; massive multiple input and multiple output (MIMO); three dimensions (3D) network architectures; femtocells & offloading solutions; green wireless technologies with energy harvesting features; and optical wireless communication (OWC).

5G comprising of ultra-dense heterogeneous networks mostly relies on revolutionary technologies such as mmW, network function virtualization, software-defined networking (SDN) and network slicing, as well as MIMO, to make a significant improvement in the transmission data rates (by $\times$ 1000), reliability, latency, and connection density (by $\times$ 1000) compared to pre-5G systems [2]. 5G is used in a wide range of applications, which can be broadly categorized into three main service classes of enhanced mobile broadband (bMBB) with data rates exceeding 1 Gbps for mobile users, ultra-reliable low-latency communications (URLLC) with high reliability (99.999%) and low latency (around 1 ms), and massive machine-type communications (mMTC) with a high number of connected devices supported in IoT deployments (up to $10^6$ devices per km$^2$).

The future 6G wireless networks should serve a range of industrial applications, such as manufacturing, healthcare, agriculture, art and culture, intelligent transportation systems, etc., and therefore must meet high requirements in terms of communication reliability ($\geq$ 99.999%), latency ($\leq$ 1ms), scalability (1 Tbps/m$^2$), energy efficiency and consider the ecosystem too [2]. This will pose new challenges to the service providers in upgrading the existing communication networks to ensure compatibility and the quality of services at low cost, which becomes highly demanding in urban areas, where front- and back-haul access networks will increase strain on the existing networks. In 6G and beyond, the integration of radio and optical wireless technologies in access networks (front- and back-haul links) will be critical in providing the requirements, particularly for bMBB, URLLC and mMTC services. Regardless of the technology (5G or 6G) being adopted, there are a few approaches to increase the capacity of wireless networks including (*i*) release of new spectrums and therefore more bandwidth, which is costly; (*ii*) using more nodes, which can be done via cell splitting, which is complex and costly. Note, that doubling the infrastructure will not lead to doubling the revenue; and (*iii*) improving the spectral efficiency, which has been done continuously over the years, but is slowing down in recent years.

Most existing wireless communication networks solely rely on the use of conventional radio frequency (RF)-based technology to convey information. However, the RF technology is currently under pressure to meet the ever-growing demand for the spectrum to cater to new application areas such as massive MIMO, machine-type smart communication for autonomous systems, augmented reality, and virtual reality. Consequently, it is imperative to investigate new materials, devices, and front-end architectures for wireless connectivity, as well as novel as well as revolutionary communication and computing paradigms. The new potential candidate technologies for 6G and beyond include **reconfigurable intelligent surfaces**, i.e., artificial planar structures with integrated electronic circuits, which can be programmed to control the incoming electromagnetic field in a wide range of functionalities, and the complimentary **optical wireless technology**.

II. OWC

Information and communication technologies contribute about 3% to the global emission of greenhouse gas, which can be significantly reduced by developing energy-efficient technology and standards [3]. Additionally, when designing the next-generation communication networks the need for network devices and equipment with lower carbon footprints must be considered. OWC can be adopted as a potential complementary technology, which will contribute to the reduced carbon footprints, that meet the demands for higher transmission throughputs in sustainable smart environments.

The OWC technologies include **free space optical (FSO)** communications (mostly infrared, but both visible and ultraviolet bands can also be used); **visible light communication (VLC),** which offers the potential of using the light emitting diode (LED)-based lighting infrastructure to establish all-optical wireless networks for data communication, positioning, and sensing mostly in indoor environments and with limited use in outdoor applications at the moment; **optical camera communication (OCC)** that can be used with both FSO and VLC systems in indoor and outdoor applications with reduced interference and high signal to noise ratio (SNR); and **OWC networking** using a combination of FSO, VLC, OCC, and RF-based schemes. Note that, indoor OWC systems are more sustainable due to their relatively lower transmit power and therefore less harmful effects compared with RF (particularly mmW) communication schemes.

OWC technology could be employed in (*a*) space for ground-to-satellite, satellite-to-ground, inter-satellite; (*b*) terrestrial (indoor and outdoor including industry, healthcare, trains. railway and bus stations, airports and aircraft, transportation, homes, offices, shopping malls,), where most Internet traffic originates and terminates in indoor environments; and (*b*) underwater, where RF technologies are not the best option. The introduction of OWC technology not only simplifies the provision of scalable, secure, and sustainable wireless systems but also facilitates the possibility of (*i*) optical positioning in indoor environments, where the use of RF-based GPS is very limited and in certain environments (tunnels, underground, etc) not possible at all; (*ii*) sensing; (*iii*) power transfer for zero IoT devices for context-aware applications; and (*iv*) releasing the RF spectrum for use in applications areas where mobility is essential and OWC cannot be employed. Compared to state-of-the-art solutions using different portions of the radio frequency spectrum (i.e., sub-6 GHz and mmW bands), OWC offers three-fold gains of:

**Ultra-densification -** Deploying an extremely large number of nodes, co-located even in the same room, without the use of advanced signal processing algorithms for interference management.

**Interaction with humans –** Using the visible light spectrum band (i.e., between 380 and 750 nm) for illuminations, sensing, and localization in indoor/underground/tunnels environments in which lights are always on, and RF-based positioning systems cannot be utilized.

**Wideband unlicensed spectrum -** Enables high point-to-point data rates when using laser-based light sources (i.e., in outdoor FSO applications), as well as high-sensing accuracy when using white LEDs for ranging and imaging. When compared to the RF signal power in the sub-mmW bands (i.e., beyond 100 GHz), strong optical power is much easier to generate in both visible and infrared bands.

Note that, in OWC the optical spectrum is much wider and is license-free, the maximum transmit power can be much higher than that allowed in RF bands, the level of interference is much lower than in RF-based systems, and the link security at the physical layer is relatively better than RF links.

In addition, hybrid wireless systems based on the integration of two or more wireless technologies, which can effectively exploit the cross-spectrum and cross-medium advantages of both OWC and mmW technologies, may offer an effective solution with enhanced features overcoming the limitations of individual wireless technologies. The hybrid wireless systems can provide seamless connectivity between fixed and wireless networks and can deliver services regardless of the fixed or mobile network. The RF/OWC hybrid system provides convergence or integration of OWC and RF networks (fixed or mobile) in several applications.

III. CHALLENGES AND FUTURE RESEARCH AREAS

Note that, while OWC offers many interesting features compared with RF technology, there are still several important and fundamental challenges and issues, see below, that need investigating to make this technology viable and applicable.

- *Channel capacity* and *channel modelling* – To develop a new theory given that optical signal must be positive and real and therefore the classic Shannonian theory is not applicable, and to better understand the wavelength-dependent reflectance and absorption of objects and materials in indoor environments.

- *Resource and spectrum utilization* – The optical spectrum is three orders of magnitude larger than the entire RF spectrum when using low bandwidth and wide emission spectrum LED light sources.

- *Link power budget and limited receiver sensitivity* – Considering the widely spread transmit optical beam (from standard LEDs) and small size photodetectors for high-speed applications, respectively.

- *Uplinks in VLC* – There is no clear idea on this and whether to use visible and/or infrared bands as well as WiFi. Consideration should be given to energy power usage since the uplink optical transmitters are very close to the users, as well as no flickering.

- *Blocking/shadowing, mobility, pointing, and tracking* - In both indoor and outdoor environments. Support horizontal and vertical handovers (in hybrid VLC-WiFi networks) to ensure seamless communications.

- *The eye and skin safety* – This is critical when using a high-power point source in indoor environments. The integration of compact and small optical concentrators in optical receivers to improve the SNR in systems using broad-beam light sources.

- *Networking and protocols* – To manage nested small cells (i.e., pico- and femto-cell) and provide seamless integration with the existing RF-based wireless and optical fiber-based backbone infrastructure. As well as the need for energy-efficient MAC protocol and green routing algorithms.

- *Link availability* – Under all-weather conditions mostly FSO systems for outdoor applications.

- *Security* – Using novel technologies such as quantum key distribution-based combined with innate physical security can result in two-fold information secrecy.

- *A dynamic architecture and network function analysis* – Where future OWC networks should have a dynamic topology in nature with a cross-layer approach because of network densification (particularly in urban areas), where the end-users will have multiple connections.

- *Resource and interference management* - Support many emerging new applications, e.g., using SDN with machine learning for real-time system management.

- *Energy harvesting* – OWC modules should be either self-powered, use ultra-long-life batteries, power over ethernet, or mains-powered, therefore is no need to keep replacing batteries.

- *Standards* – More contributions from academia and industry to the existing standards on short-to-long range OWC systems.

- *Optical beamforming/beam steering optimization using machine learning* – This facilitated allocating the optical radiation to specific users, thus improving SNR and security as well as reducing interference.

- *New devices* – Investigation of new devices with improved optical and electrical characteristics that will meet the energy, bandwidth, sustainability, and adaptability requirements of the next-generation wireless networks.

- *Simple and low-cost plug-and-use modules* – This is most important in VLC and OCC systems if they are going to be widely adopted by the industry and the end-users.

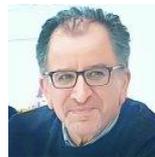

**Professor Zabih Ghassemlooy**, Fellow, SOA; Fellow, IET; Senior Member, IEEE; Member, ACM), CEng, BSc (Hons.) in EEE, Manchester Metropolitan Univ., (1981), MSc (1984) and PhD (1987) from Manchester Univ., UK. 1987-88 Post-Doctoral Research Fellow, City Univ., UK. 1988-2004 Sheffield Hallam University, UK, and 2004-14 joined Faculty of Eng. & Env., Northumbria University, UK as an Associate Dean Research, and currently is the Head of Optical Communications Research Group. He is a Research Fellow (2016-) and a Distinguished Professor (2015-) at the Chinese Academy of Science. He was the Vice-Chair of EU Cost Action IC1101 (2011-16) and is Vice-Chair of the EU COST Action CA19111 NEWFOCUS (European Network on Future Generation Optical Wireless Communication Technologies, 2020-2024). Published 980 papers (415 journals and 8 books), over 110 keynote/invited talks, supervised 12 Research Fellows and 73 PhD students. Research interests are in areas of optical wireless communications (OWC), free space optics, visible light communications, hybrid RF-OWC, software-defined networks with funding from EU, UK Research Council, and industries. He is the Chief Editor of the British J. of Applied Science and Technology and the International J. of Optics and Applications, Associate Editor of several international journals, and Co-guest Editor of several special issues OWC. He is the Vice-Cahir of OSA Technical Group of Optics in Digital Systems (2018-); the Chair of the IEEE Student Branch at Northumbria University, Newcastle (2019-). 2004-06 was the IEEE UK/IR Communications Chapter Secretary, the Vice-Chairman (2006-2008), the Chairman (2008-2011), and Chairman of the IET Northumbria Network (Oct 2011-2015). z.ghassemlooy@northumbria.ac.uk.

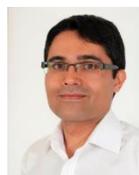

**Dr Mohammad Ali Khalighi** (Senior Member, IEEE) is Associate Professor with École Centrale Marseille, Marseille, France, and head of "Optical Communications for IoT" group at Fresnel Institute research lab. He is currently serving as Action Chair for the COST Action CA19111 NEWFOCUS (European Network on

Future Generation Optical Wireless Communication Technologies). He was the Coordinator of the H2020 ITN MSCA VisIoN project (Visible-light-based Interoperability and Networking). He has coedited the book "Visible Light Communications: Theory and Applications" (CRC Press, 2017) and was the co-recipient of the 2019 Best Survey Paper Award of the IEEE Communications Society. He is serving as Editor-at-Large for the IEEE Transactions on Communications, and also served as Associate Editor for the IET Electronics Letters as well as Lead Guest Editor for the IEEE Open Journal of the Communications Society and Elsevier Optik journal. His main research interests include wireless communication systems with an emphasis on free-space, underwater, and visible-light optical communications. Ali.Khalighi@fresnel.fr.

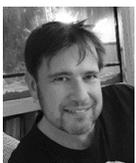

**Professor Stanislav Zvanovec,** Senior Member, IEEE, received the M.Sc. and Ph.D. degrees from the Faculty of Electrical Engineering, Czech Technical University (CTU) in Prague, in 2002 and 2006, respectively. He is currently works as a Full Professor, the Deputy Head of the Department of Electromagnetic Field, and the Chairperson of Ph.D. Branch with CTU. He leads Wireless and Fiber Optics team (optics.elmg.org). His current research interests include free space optical and fiber optical systems, visible light communications, OLED, RF over optics, and electromagnetic wave propagation issues for millimeter wave band. He is the author of two books (and coauthor of the recent book Visible Light Communications: Theory and Applications), several book chapters, and more than 300 journal articles and conference papers. xzvanove@fel.cvut.cz.

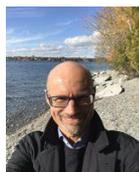

**Prof Nobby Stevens** (M'14) received the master's degree in physical engineering from Ghent University, Ghent, Belgium, in 1997, the D.E.A. degree from the Institut National Polytechnique de Grenoble, Grenoble, France, in 1997, and the Ph.D. degree from Ghent University in 2004. From 1997 to 1998, he was a Product Development Engineer with Philips. Beginning in 1998, he was with the Department of Information Technology, Ghent University, where he performed research on numerical modeling of electromagnetic fields interacting with the human body. In 2004, he joined Agilent EEsof, Santa Rosa, CA, USA, as a Research and Development Engineer involved in computational electromagnetics. Since 2008, he has been performed research with the DraMCo (wireless and mobile communications) Group, ESAT, KU Leuven, Ghent, where he is associate professor since 2018 . His research activities are mainly focused on optical wireless communications, with a strong emphasis on the deployment of indoor positioning. nobby.stevens@kuleuven.be.

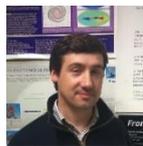

**Dr Luis Nero Alves** IEEE member since 2003. Graduated in 1996 and received his MSc degree in 2000, both in Electronics and Telecommunication Engineering from the University of Aveiro. Obtained the PhD degree in Electrical Engineering from the University of Aveiro in 2008. His expertise is on integrated circuit design for optical wireless applications. Since 2008 he has been the lead researcher from the Integrated Circuits group at Instituto de Telecomunicações, Aveiro. His current research interest are associated to the design and performance analysis of optical wireless communication systems, with special flavor in visible light communications. Other research topics include materials and devices for IoT sensing devices. He has authored and coauthored several research articles in these fields. Luis Nero Alves has participated/coordinated several research projects in the fields of optical wireless communications and sensing devices (with both national funding – FCT/VIDAS, FCT/EECCO, IT/VLCLighting - and international funding – EU-CIP/LITES, EU-FP7/RTMGear, EU-COST/OPTICWISE, EU-COST/MEMOCIS, EU-/H2020/MSCA/VisIoN, EU-H2020/FET/NeuralStimSpinal). Luis Nero Alves has also served in the technical program committee of several international conferences on the field of optical wireless communications. He has also served as reviewer for several journals in Optical communications, amongst which, IEEE/PTL, IEEE-OSA/JLT and Elsevier/OC.
nero@ua.pt.

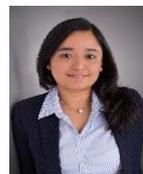

**Amita Shrestha** from German Aerospace Center (DLR), received her bachelor's degree in Electronics Engineering and communications from Kathmandu University, Nepal in 2006. She received her masters' degree in Communications Systems and Electronics form Jacobs University, Bremen in 2009. Since 2010 she is working in the Institute of Communications and Navigation in DLR focusing on Free space Optical Communications. In DLR, she has been involved in the development of real-time tracking software of Institute's optical ground stations, and its operation during several satellite and aircraft downlink experiments like OPALS, SOTA, DoDfast, VABENE etc. Currently, she is actively involved in the standardization of optical links in CCSDS (The Consultative Committee for Space Data Systems) community and several other projects related to free space optical classical and quantum communications. Additionally, she is chairing one of the working groups in COST Action CA19111 NEWFOCUS (European Network on Future Generation Optical Wireless Communication Technologies, 2020-2024) focusing on long range free space optical communication links. amita.shrestha@dlr.de.

# 3. VISIBLE LIGHT COMMUNICATIONS FOR INTELLIGENT TRANSPORTATION SYSTEMS

Murat Uysal
Ozyegin University, Turkey, murat.uysal@ozyegin.edu.tr

## I. INTRODUCTION

Intelligent Transportation Systems (ITSs) are built upon cooperation, connectivity, and automation of vehicles, and are expected to improve the safety, efficiency, and sustainability of passengers and freight while enhancing the comfort of driving. The practical implementation of ITSs requires highly reliable, robust, and scalable vehicle-to-everything (V2X) communication solutions enabling vehicle-to-infrastructure (V2I) and vehicle-to-vehicle (V2V) connectivity. The information gathered via V2X communications provides the drivers with real-time information on traffic and road conditions including collisions, congestion, surface condition, traffic signal violations, emergency brakes, etc. Furthermore, such information can be used by the local transportation departments to develop traffic efficiency applications, e.g., platooning, or by automotive manufacturers to develop safety applications, e.g., collision avoidance. Besides these road safety functionalities, potential in-vehicle applications have recently emerged such as high-speed internet access, multiplayer gaming, cooperative downloading, and mobile commerce because of ever-increasing dependency on the Internet and multimedia services.

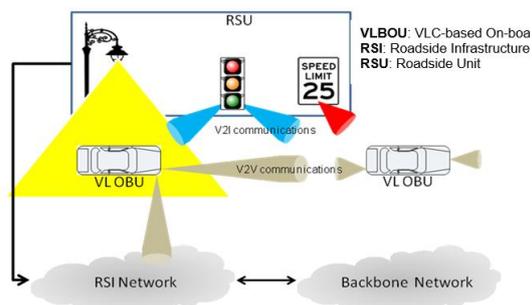

**Fig. 1.** Vehicular VLC network

Research efforts and standardization activities on V2X have been so far centered around radio frequency (RF) technologies. In 2010, as an amendment of the IEEE 802.11 WiFi standard, IEEE 802.11p standard was introduced to support V2X communications in the 5.9 GHz band allocated for ITS applications. In 2017, a cellular-based V2V connectivity solution, known as LTE-V, was introduced as a part of 3GPP Release 14. Since the current market penetration of V2X solutions is relatively low, the allocated RF bands are considered sufficient at the time. However, soon, high interference levels can be experienced in limited RF bands, particularly in high-density traffic scenarios. Channel congestion will result in longer delays and degrade the packet rate. To address such issues, visible light communications (VLC) has been proposed as an alternative vehicular access solution to RF-based V2X communications [1]. VLC takes advantage of widely available light emitting diodes (LEDs) and utilizes them as wireless transmitters in addition to their primary illumination function. Since automotive manufacturers are increasingly using LED-based exterior lighting, front and back vehicle lights can serve as wireless transmitters making VLC a natural vehicular connectivity solution.

## II. VLC-BASED V2X: BASIC CONCEPT AND ADVANTAGES

As illustrated in Figure 1, a vehicular VLC network consists of onboard units (OBUs), i.e., vehicles, and roadside units (RSU), i.e., traffic lights, street lamps, digital signage etc. Additionally, RSUs are connected to the backbone network via the Road Side Infrastructure (RSI) network. Vehicles fitted with LED-based front and back lights can communicate with each other and with the RSUs through the VLC technology. Furthermore, LED-based RSUs can be used for both signaling and broadcasting safety-related information to vehicles on the road.

In comparison to RF counterparts, VLC offers inherent advantages such as immunity to the electromagnetic interference, operation in unlicensed bands, inherent security, and a high degree of spatial confinement allowing a high reuse factor. VLC is well-positioned to address both the low latency required in safety functionalities (i.e., emergency electronic brake lights, intersection collision warning, in-vehicle signage, platooning) and high speeds required in so-called infotainment applications (i.e., map downloads and updates, media downloading, point of interest notification, media downloading, high-speed internet access, multiplayer gaming, etc). Furthermore, VLC is a cost-effective and green communication solution since the dual use of LED lighting systems on vehicles and the roadside infrastructure is targeted. A LED-based VLC system would consume less energy compared to the RF technology, thus allowing the expansion of communication networks without added energy requirements, potentially contributing to the global carbon emissions reduction in the long run. VLC is also appealing for scenarios in which the use of RF band is restricted or banned due to the safety regulations, e.g., industrial parks such as in oil/gas/mining industries. VLC could be also used for positioning and navigation purposes. Although GPS is widely used today, it fails to provide sufficient accuracy in environments where there is no line-of-sight (LOS) path such as tunnels, indoor parking lots, and some urban canyons. For such cases, VLC-based positioning technology is more attractive.

III. STATE-OF-THE-ART

With attractive features and potential application areas discussed within the previous section, vehicular VLC has received increasing attention lately, see the recent surveys in [2] and [3]. The current literature can be roughly divided into two categories based on the type of receptor. A vehicular VLC system can deploy a photodetector or an image sensor (camera) to receive the optical signal transmitted by the LED-based head- and taillights. Since cameras are already deployed in most vehicles for safety applications such as parking assist and lane detection, these built-in cameras can be also potentially used for VLC systems. In such a camera-based VLC system, the received light from the imaging lens projected onto the image sensor is converted to binary data by the readout circuit. The image sensor consists of multiple micron- sized reception pixels. Based on the configuration of the readout circuit, image sensors use either rolling shutter or global shutter technology. In the latter typically used with CCD image sensors, all pixels are read at once. In the rolling shutter technology typically preferred in CMOS image sensors, pixels are read one row/column at a time, which makes them relatively faster in comparison to the global shutter. Nevertheless, their frame rate (typically 30–100 fps) results in a throughput of tens of bits per second limiting their application to mainly basic safety functionalities. The limitations of camera-based solutions have prompted researchers to explore the use of photodetectors to enable much higher data speeds in vehicular VLC systems. PIN (p-n diode) photodetectors are more favorable for vehicular VLC due to lower cost as well as better linearity performance and high-temperature tolerance. Despite a higher cost, avalanche photodiode (APDs) have higher sensitivity and provide better gain compared to PIN counterparts which make them a more robust solution, particularly in adverse weather conditions.

The current literature on vehicular VLC is well established in terms of channel modeling and physical layer design. Different from indoor LEDs, the headlights and taillights exhibit asymmetrical intensity distribution. Considering such vehicular LED characteristics as well as weather conditions, channel models for vehicular VLC were developed and extensive simulation studies and analytical results were presented to demonstrate the fundamental limits of vehicular VLC systems, see e.g. [4]. While earlier works on vehicular VLC are limited to simple pulse modulation techniques and single-hop configurations, more recent works have demonstrated significant improvements via the use of more sophisticated physical layers such as optical orthogonal frequency division multiplexing (O-OFDM) and its variants. O-OFDM was shown to be effective in handling the intersymbol interference resulting from multipath propagation and the limited bandwidth of LED. Multi-hop transmission techniques made it possible for the signal transmitted from the source vehicle to reach the destination vehicle through a number of intermediate vehicles (relays) eliminating the need for LOS requirements. Hybrid VLC/RF links were further proposed to ensure link availability under all weather conditions.

IV. CHALLENGES AND FUTURE DIRECTIONS

While recent experimental works have already demonstrated the feasibility of the vehicular VLC, further research efforts are required in several areas of vehicular VLC before commercialization and widespread adoption of this promising technology. In particular, the dynamic conditions imposed by the outdoor medium (i.e., adverse weather, the effect of sunlight etc.) and vehicle mobility necessitate the design of adaptive transmission solutions. At the physical layer, link adaptation might involve the selection of modulation type/size, channel code rate, and/or transmit power based on the instantaneously received signal-to-noise ratio. On the hardware level, the vehicular VLC system can be modified to enable dynamic adaptation of its receiver field-of-view (FOV) or use an adjustable optical attenuator to minimize the incoming background noise.

In order to transform vehicular VLC into a full-fledged solution, additional efforts in the upper layers are further required. For example, most of the medium access control (MAC) protocols in the literature assume isotropic radiation of RF systems. The fact that VLC systems with their inherent directionality render conventional schemes practically useless dictates the need for the design of novel MAC protocols that consider the directionality of the illumination pattern of headlights and taillights. Another critical research topic that requires further investigation is the integration and co-existence of vehicular VLC with RF-based technologies such as IEEE 802.11p and C-V2X. Initial experimental results have shown that such heterogeneous solutions can compensate for the drawbacks of each other and improve the overall performance. However, additional efforts are required for a full integration at the hardware level possibly exploiting the common system architecture based on OFDM.

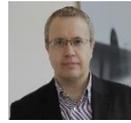

Prof Murat Uysal received the B.Sc. and the M.Sc. degree in electronics and communication engineering from Istanbul Technical University in 1995 and 1998, respectively, and the Ph.D. degree in electrical engineering from Texas A&M University in 2001. He is currently a Full Professor of the Department of Electrical and Electronics Engineering at Ozyegin University, Istanbul, Turkey. He also serves as the Founding Director of the Center of Excellence in Optical Wireless Communication Technologies (OKATEM). Prior to joining Ozyegin University, he was a tenured Associate Professor at the University of Waterloo, Canada. Dr. Uysal's research interests are in the broad area of communication theory with a particular emphasis on the physical layer aspects of wireless communication systems in radio and optical frequency bands. He has authored some 400 journal and conference papers on these topics and received more than 17.000 citations with an h-index of 61. Dr. Uysal is an IEEE Fellow and a member of Turkish Science Academy. He currently serves as an Area Editor for IEEE Transactions on Communications. In the past, he served as an Editor for IEEE Transactions on Wireless Communications, IEEE Transactions on Communications, IEEE Transactions on Vehicular Technology, and IEEE Communications Letters. Dr. Uysal's major distinctions include the NSERC Discovery Accelerator Award, University of Waterloo Engineering Research Excellence Award, Turkish Academy of Sciences Distinguished Young Scientist Award, Ozyegin University Best Researcher Award, National Instruments Engineering Impact Award and IEEE Turkey Section Outstanding Service Award in addition to numerous best paper awards.

# 4. ON THE USE OF VISIBLE LIGHT COMMUNICATIONS IN INTELLIGENT TRANSPORT SYSTEMS: FROM ROAD VEHICLES TO TRAINS

Anna Maria Vegni

Department of Industrial, Electrical and Mechanical Engineering, Roma Tre University, Italy,
nnamaria.vegni@uniroma3.it

## I. INTRODUCTION

Visible light communication (VLC) technology is foreseen as a key technology in the context of next generation 6G wireless networks, mainly due to its spectral, energy, and cost efficiencies. VLC is expected to be adopted as a complementary technology for radio frequency (RF) networks, especially in places where RF networks fail to support end-users with their quality-of-service requirements. Furthermore, VLC is a candidate as a green technology, providing communication enhancements due to high data rates, while guaranteeing illumination and security requirements.

The dual paradigm of providing both illumination and communications allows VLC to be suitable for several application scenarios, where lighting devices are used. Ranging from vehicle-to-vehicle (V2V) and vehicle-to-infrastructure (V2I) communications to multi-user data offloading, as well as high-speed train communications, the use of VLC systems can be largely adopted in Intelligent Transport Systems (ITS), mainly for safety and positioning applications. VLC has been initially identified to support safety applications in vehicular ad-hoc networks, considering packet transmission from a vehicle to the forwarding one, by means of multiple hop V2V modes. Also, the use of fixed nodes, like Roadside Units, and the existing infrastructure, has allowed packet transmission according to V2I-I2V mode.

This paper presents a brief description of the use of VLC in ITS, in the case of different application domains, ranging from vehicular networks to railways, as presented in Section 2. In particular, VLC is envisioned as a promising technology in railway, due to the need of overcoming existing technology solutions that are becoming obsolete, such as the GSM-R standard. As of today, GSM-R is going to be substituted by novel solutions for managing wireless multi-bearer links for train-to-ground (and vice versa) available along the rail path, such as Future Railway Mobile Communication System and Adaptable Communication System. A short discussion about this aspect is finally provided at the end of this paper.

## II. RELATED WORK

The main vehicle safety applications are identified by analyzing the occurring frequency and the impact of different classes of accidents. The most representative safety applications are summarized in Table 1, showing the high-priority applications with very strict limits concerning the latencies [1], with values below 100 and 20 ms limits for the pre-crash sensing, and a maximum required communication range that varies between 50 and 300 m. Information about communication mode and message length are also provided. This points out that in vehicular applications, up to a certain limit, the connectivity, the robustness, and the latencies are prior to communication distances.

In ITS, VLC considers LEDs as transmitting devices and photodetector or image sensors as receivers. In the latter case, image sensors are adopted in Optical Camera Communications, which represents a sub-class of OWC systems. In OCC systems for ITS applications, image sensors are mounted both on the taillights and on the front side of every vehicle. Each vehicle broadcasts its own ID and other information to neighboring vehicles. Images are captured and detected using deep learning techniques among the numerous interfering light sources on the road. In [2], the schematic of image capture from the LED taillights, by means of convolutional neural networks, is presented, to detect the distance between two neighboring vehicles.

In railway scenarios, VLC applicability mainly refers to data communications, and voice and video-streaming services. The deployment of VLC systems can be easily adopted, by replacing existing luminaries used for illumination on the trains with white LEDs. However, the applicability of VLC systems can be adopted not only on the train itself but also on station buildings, along trackside, inside tunnels, as well as in case of station/yard scenarios [3].

In the case of a mainline/highspeed scenario, where train speed can reach very high values, VLC systems can be exploited for data communications and signaling, related to the train operations. In [4], it was described as the signal passed at danger (SPAD) risk, which refers to the occurrence of a train crossing the stop signal without the authority to do so, thus causing accidents. SPAD can be caused by several factors, including driver's inattention, distraction, fatigue, incomplete route knowledge, misunderstanding, and poor sighting of trackside signals. In this situation, VLC can be applied to increase the efficiency of the current methods of avoiding SPADs, such as the Automatic Warning System and Train Protection & Warning System, by means of their integration. Specifically, the emitting lights of the trackside signal can be detected by the photodetectors laying on the train, and then the information will be decoded as a visual/ audio alert shown on the monitor of the train driver.

| Application | Maximum range [m] | Maximum latency [ms] | Message length [bit] | Type |
|---|---|---|---|---|
| Traffic Signal Violation Warning | 250 | 100 | 528 | I2V |
| Curve Speed Warning | 200 | 100 | 235 | I2V |
| Emergency Electronic Brake Light | 300 | 100 | 288 | V2V |
| Pre-crash Sensing for Cooperative Collision Mitigation | 50 | 20 | 435 | V2V |
| Cooperative Forward Collision Warning | 150 | 100 | 419 | V2V |
| Left Turn Assistant | 300 | 100 | 904, 208 | I2V / V2I |
| Lane Change Warning | 150 | 100 | 288 | V2V |
| Stop Sign Movement Assistant | 300 | 100 | 208, 416 | V2V / I2V |

**Table 1**. Classification of high-priority safety ITS (vehicular) applications, [1]

III. CHALLENGES AND CONCLUSIONS

As emerged, the use of VLC systems in vehicular networks is consolidated, while more recently in railway applications it is largely increasing, also due to the pressing request for enhanced performance of new rail applications. H2020 AB4Rail project (https://www.ab4rail.eu/) aims to identify novel emerging technologies, eligible to replace GSM-R and allow new railway services and applications. It has been envisaged the usage of aerial and space technologies, such as HAPS and high-throughput LEO satellites, as well as optical wireless technologies like VLC and Free Space Optics. For each alternative bearer, different technological features and limitations are being investigated in AB4Rail project, and it is expected that the use of synergy among multiple bearers can be suitable to guarantee performance in different railway scenarios and applications.

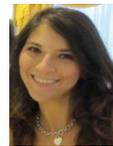
**Dr Anna Maria Vegni** (Senior member, IEEE) is a tenure-track Assistant Professor in the Department of Engineering at Roma Tre University (Rome, Italy), since March 2020. She received the Ph.D. degree in Biomedical Engineering, Electromagnetics and Telecommunications from the Department of Applied Electronics, Roma Tre University, in March 2010. She received the 1st and 2nd level Laurea Degree cum laude in Electronics Engineering at Roma Tre University, in July 2004, and 2006, respectively. In 2009, she was a visiting researcher in the Multimedia Communication Laboratory, directed by Prof. Thomas D.C. Little, at the Department of Electrical and Computer Engineering, Boston University, Boston, MA. Her research activity focuses on vehicular networking, optical wireless communications, and visible light positioning. She a member of ACM and an IEEE Senior Member. In June 2021, she got the Italian Habilitation (Abilitazione Scientifica Nazionale) for Full Professorship in Telecommunication Engineering. She is involved in the organization of several IEEE and ACM international conferences and is a member of the editorial board of IEEE TCOM, IEEE ComMag, Ad Hoc Networks, Journal of Networks and Computer Applications, Nanocomnet Elsevier journals, WINET Springer, IEEE JCN, ITU J-FET and ETT Wiley journal.

# 5. UNMANNED AERIAL VEHICLES WITH LIGHTWAVE TECHNOLOGY FOR 6G NETWORKS

Panagiotis D. Diamantoulakis, Vasilis K. Papanikolaou, and George K. Karagiannidis
Wireless Communications & Information Processing Group, Department of Electrical and Computer Engineering, Aristotle University of Thessaloniki, 54 124, Thessaloniki, Greece,
{padiaman, vpapanikk, geokarag}@auth.gr

## I. INTRODUCTION

The integration of terrestrial and aerial networks is one of the key objectives of wireless networks beyond the fifth generation (5G). Toward this direction, the unmanned aerial vehicles (UAVs) are envisioned to be used as part of the network's infrastructure for coverage extension, traffic offloading in crowded environments, and rapid recovery of the network services in cases of emergency. More specifically, the use of UAVs as ad-hoc mobile base stations can have several important real-world applications, such as vehicular networks and train backhauling, building/human health monitoring, precision agriculture, the interconnection of critical infrastructure such as smart grids in a secure way, and virtual reality. Also, UAVs can be utilized as mobile edge servers in order to support functionalities such as mobile edge computing and federated learning, with the potential to reduce delay and improve privacy. The latter is of paramount importance for 6G networks, which can be seen as the superposition of communication and computing networks, in which distributed nodes are able to support artificial intelligence applications [1].

In the integration of UAVs and ground networks, energy efficiency, cost, and ease of deployment need to be considered, which also constitute important challenges toward the 6G networks. The effective use of UAVs depends on the communication performance and the UAVs' flight time duration. The bottleneck of the communication performance is usually imposed by the front hauling link, i.e., the ground-to-air (G2A) link. On the other hand, the flight time of the UAVs is limited by their battery capacity, while, in the general case, the UAVs' power consumption is composed of the propulsion power, the communication-related power, e.g., for the communication between the UAVs and the users via the air-to-ground (A2G) links, and the power consumption for data processing. To address the aforementioned challenges, the use of hybrid optical wireless communication (OWC)/radio frequency (RF) technology is particularly promising.

## II. THE CONCEPT OF ENERGY AUTONOMOUS UAV-BASED COMMUNICATIONS SYSTEMS WITH SLIPT

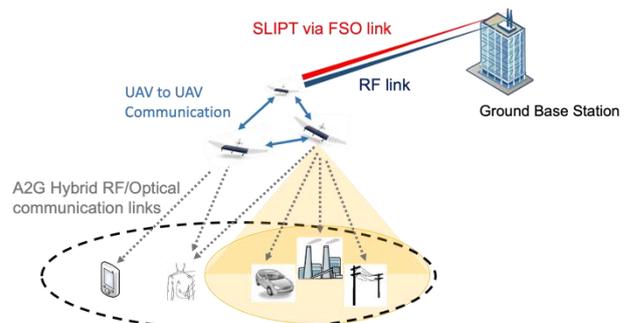

**Fig. 1.** Cooperative air-to-ground and ground-to-air hybrid RF/optical system with simultaneous lightwave information and power transfer (SLIPT)

In Fig. 1, the reference architecture of a multi-UAV cooperative G2A-A2G network is illustrated. In the proposed setup, multiple UAVs are used to serve multiple users in an ad-hoc way. To mitigate the front hauling bottleneck, a dedicated free space optics (FSO) link can be used for the communication between the ground base station and the UAVs, as rates up to 80Gbps have been reported in the literature for FSO G2A communications, while in order to further increase the data rates and especially reliability, an RF link, e.g., mmWave, can be used in a complementary way. The communication between the UAVs and the users can also be based on hybrid RF/OWC technology, while the directivity of the corresponding OWC links depends on the use case. For example, FSO may be the best option for the point-to-point communication between UAVs and smart grids infrastructure, while visible light communication (VLC) or infrared (IR) based broadcasting technology can be utilized for the information transmission between UAVs and multiple vehicles. Also, when using the RF technology, except for increased reliability and throughput when used in parallel with the OWC links, it is necessary to retain connectivity with conventional mobile users and Internet-of-Things sensors. Furthermore, RF can be used for the exchange of control signals among UAVs, which can collaborate to perform swarm operations. Finally, in order to meet the energy requirements of the UAVs and extend their flight lifetime, energy can be transferred via the dedicated FSO link by using simultaneous light wave information and power transfer (SLIPT) technology.

## II. STATE OF THE ART

As it has been proposed in [2], SLIPT can be implemented by equipping UAVs with solar panels. In this regard, by using lightwave power transfer, flight time has been shown to be increased up to 24 times, while other reports showcase a 200W power transfer for a distance up to 100m. To explore the potential of using SLIPT as an enabler of energy-autonomous UAV-based networks, relevant research attempts have mainly focused on three different directions, namely the optimization of SLIPT, channel modelling, and the investigation of the impact of network topology on the achieved performance. In more detail, the use of SLIPT creates an interesting trade-off between the harvested energy and communication performance of the FSO link, calling for the optimization of the resources, e.g., optical power and time, that are allocated to energy transfer and information transmission [2]. However, although the analysis in [2] provides the optimal configuration with respect to the instantaneous channel gains, to characterize the long-term performance of UAV-based FSO systems, appropriate theoretical channel models are needed.

Conventional FSO channel models are not applicable in UAV-based communications, which is mainly due to the UAVs' mobility and the random fluctuations of UAVs' position and orientation because of the dynamic wind load, and random air fluctuations in the atmosphere around the UAVs, and the internal vibrations of the UAVs. Thus, the incident laser beam is in the general case non-orthogonal to the receiver plane, while the fluctuations may be correlated [2]. To this end, theoretical models for the geometrical and misalignment losses caused by the random fluctuation of the position and orientation of the UAVs have been proposed in [3]. Also, in practice, the flight time of UAVs is interdependent on the network's topology and consistency, which can be modelled by using stochastic geometry-based tools [4].

## III. CHALLENGES & FUTURE RESEARCH DIRECTIONS

### A. Experimental Channel Modelling

As it has already been mentioned, the precise statistical modelling of channel is of paramount importance for the quantification of the stability requirements of UAVs in order to maintain a certain link quality, which is determined by the communication quality-of-service requirements, as well as the required amount of harvested energy when the G2A FSO link is used for SLIPT. However, to fully address the aforementioned issues, experimental results are needed, as well as the matching of the statistical parameters with the UAVs' specifications. Similarly, another challenging issue is the statistical modelling of the VLC/IR G2A links.

### B. Cross-layer Optimization

For the design, orchestration, and resource allocation of UAV-based networks of Fig. 1, several different parameters need to be jointly optimized, such as the size and weight of the UAVs, the capacity of the batteries, the size of the information buffers, the transceivers design and the photodetector's light collecting area, the UAVs trajectory or placement, the utilized PHY and MAC layers techniques, the tracking system, the handover mechanism, the data-processing requirements, etc. In addition, the use of multiple frequency bands has the potential to increase the data rates, but it also increases complexity and, thus, the corresponding energy consumption. These considerations create several interesting trade-offs that require cross-layer optimization approaches. The formulated problems may be particularly challenging, making the use of model-free machine learning methods in order to solve them a promising alternative.

### C. Quantum Key Distribution through UAVs

One of the most promising uses of UAV-enabled FSO communications is to increase communication security in critical applications. Current security protocols are limited by their public use of unsecured communication channels and as such, they are vulnerable to computational and hardware advancements. To address that, quantum communications (QC), that are facilitated through the reliable transmission of quantum states, have given rise to quantum key distribution (QKD). QKD offers superior efficiency in securing key distribution by capitalizing on the laws of quantum mechanics while it has already been implemented in both fiber optic networks and through free space optics for terrestrial and aerial links. However, in optical fibers, path loss scales exponentially with distance, whereas in atmospheric links, such as UAV-to-ground links, the main cause of loss is diffraction, which scales quadratically with distance. This condition makes FSO through aerial platforms a prime candidate to implement QKD. There are two main ways that QKD can be implemented. The most common is to encode the information onto a quantum state of a single photon, such as phase or polarization. Although single-photon receivers can be developed, they are costly and highly cumbersome to be maintained on a UAV. On the other hand, more recently, the continuous variable QKD has been implemented by encoding the key onto the continuous quantum variables. The main challenge for FSO QKD is that accurate channel modeling for atmospheric QC links is required since weather effects and pointing errors due to misalignment and jitter can severely inhibit the transmission.

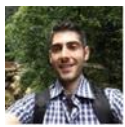
**Dr Panagiotis D. Diamantoulakis** received the Diploma (5 years) and PhD from the Department of Electrical and Computer Engineering (ECE), Aristotle University of Thessaloniki (AUTH), Greece, in 2012 and 2017, respectively. Since 2017, he works as a Post-doctoral Fellow in Wireless Communications & Information Processing (WCIP) Group at AUTH and since 2021, he is a visiting Assistant Professor in the Key Lab of Information Coding and Transmission at Southwest Jiaotong University (SWJTU), China. His research interests include optimization theory and applications in integrated communication and computing networks, optical wireless communications (OWC) and hybrid OWC/RF networks, and wireless power transfer. He is a Working Group Member in the Newfocus COST Action "European Network on Future Generation Optical Wireless Communication Technologies". He is Senior IEEE Member. He serves as an Editor for IEEE Wireless Communications Letters, IEEE Open Journal of the Communications Society, Physical Communications (Elsevier), and Frontiers in Communications and Networks. He was also an Exemplary Editor of IEEE Wireless Communications Letters in 2020, an Exemplary Reviewer of IEEE Communications Letters in 2014 and IEEE Transactions on Communications in 2017 and 2019.

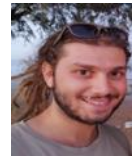
**Vasilis K. Papanikolaou** was born in Kavala, Greece in 1995. He received the Diploma Degree (5 years) in Electrical and Computer Engineering from the Aristotle University of Thessaloniki (AUTH), Greece, in 2018, where is currently pursuing his PhD with the Department of Electrical and Computer Engineering. He was a visitor researcher at Lancaster University, UK, at Khalifa University, Abu Dhabi, UAE, and at Northumbria University, Newcastle upon Tyne, UK. In 2018, he received the IEEE Student Travel Grant Award for IEEE WCNC 2018. His research interests include optical wireless communications (OWC), non-orthogonal multiple access (NOMA), optimization theory, and game theory. He has served as a reviewer in various IEEE journals and conferences. He was also an Exemplary Reviewer of IEEE Wireless Communications Letters in 2019 (top 3% of reviewers).

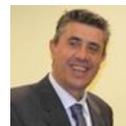
**Prof George K. Karagiannidis** received the University Diploma (5 years) and PhD degree, both in electrical and computer engineering from the University of Patras, in 1987 and 1999, respectively. He is currently Professor in the Electrical & Computer Engineering Dept. and Head of Wireless Communications & Information Processing (WCIP) Group. He is also Honorary Professor at South West Jiaotong University, Chengdu, China. His research interests are in the broad area of Digital Communications Systems and Signal processing. He is IEEE Fellow. In the past, he was Editor-in-Chief of IEEE Communications Letters from 2012 to 2015 and Associate Editor-in-Chief of IEEE Open Journal of Communications Society from 2019 to 2022. He is one of the highly-cited authors across all areas of Electrical Engineering, recognized from Clarivate Analytics as Web-of-Science Highly-Cited Researcher in the last seven consecutive years 2015-2021. Prof. Karagiannidis received the 2021 IEEE Communications Society Radio Communications Committee Technical Recognition Award and the 2018 Signal Processing and Communications Electronics Technical Recognition Award of the IEEE Communications Society.

# 6. OPTICAL WIRELESS COMMUNICATIONS TECHNOLOGIES FOR SMART MANUFACTURING IN INDUSTRY 4.0

Beatriz Ortega and Vicenç Almenar

Instituto de Telecomunicaciones y Aplicaciones Multimedia, Universitat Politècnica de València, Camino de Vera, s/n 46022 Valencia (SPAIN), {bortega; valmenar}@dcom.upv.es.

## I. INTRODUCTION

Over the constant evolution of industries towards further improvements of efficiency and quality of the products, current industrial generation is commonly referred as Industry 4.0. In this paradigm, the connection of the production systems to a communication network, called cyber-physical system, becomes critical to allow industrial automation and smart manufacturing, where Software-Defined Cloud Manufacturing and augmented reality can be introduced into intelligent production systems. 5G communication networks provide reliable links with high capacity, low latency, and low jitter between machines, sensors, and computing systems but pioneering works are being conducted toward next-generation wireless networks. More concretely, the International Telecommunication Union (ITU-T) standardization sector launched a focus group called Technologies for Network 2030, which is expected to consider new potential use cases such as holographic-type communications, ubiquitous intelligence, Tactile Internet, multi-sensor experience and digital twin in smart industrial environments, see Fig. 1 [1].

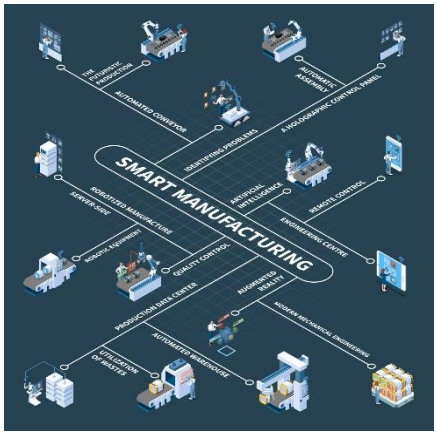

**Fig. 1.** Examples of use cases in future smart manufacturing (freepik.es)

6G new scenarios have been proposed from 5G ITU-R M.2083 definitions to meet the requirement of such use cases including ubiquitous mobile broad-band (uMBB), ultra-reliable low-latency broadband communications (ULBC) and massive ultra-reliable low-latency communications (mULC) [1]. Different applications including machine motion control, mobile robots or mobile cranes amongst others were listed by 3GPP TR 22.804 and TS 122.104 with corresponding requirements, which can be summarised as several Gbps data rates, high demand for reliability ($> 99.99\%$) and low cycle times (tens of μs).

Such increase in transmission data rates, the density of connections, and latency requirements lead to spectrum exhaustion and traffic congestion which can be overcome by using higher radio frequencies (millimeter wave and even THz range) at the expense of limited reach. An attractive solution is provided by optical wireless networking which exploits 320 THz bandwidth in the visible light spectrum (400-700 nm) and/or 12.5 THz in the infrared spectrum centered at 1550 nm and has been identified as the main enabler due to the robustness against the electromagnetic interference, availability of unregulated optical frequencies, high reliability, and high security (i.e., spatial confinement). Moreover, due to the aforementioned advantages, OWC is also proposed to integrate high-accuracy positioning systems which are also essential in industrial environments for smart manufacturing [2], as explained below.

## II. STATE OF THE ART

However, the most significant limitation of OWC systems in smart factories is the blockage of the line-of-sight (non LOS), i.e., due to moving robots. Current solutions utilize relatively wide beams, antenna diversity or MIMO schemes. In [3] a distributed multiuser MIMO is proposed for industrial plants to cope with the problems associated with blockage situations. A dense network of optical frontends (OFE) transmitters coordinated by a unique access point (AP) allows to implement spatial diversity or spatial multiplexing and supports mobility along the hall as the AP dynamically selects the set of OFE that covers the user's position. Several beam steering methods in OWC systems have been previously reported by using tunable MEMS and spatial light modulators (SLM) and, also, by using passive elements such as a pair of gratings for 2D steering or an arrayed waveguide grating router. In [4] narrow 2D steered infrared beams were proposed to be employed in Industry 4.0 to create separate and close links to each device where a matrix of photodiodes allows the deployment of a wide field of view receiver to facilitate the alignment in smart factories.

Moreover, other approaches are based on the concept of intelligent reconfigurable reflecting surfaces and the use of diffuse reflection focusing on beam reconfiguration has been recently proposed in order to simplify the indoor environments of Industry 4.0. In this system, the light beam is reflected in a small area with waveform shaping based on a SLM, focused, and steered to the desired user by using a transfer matrix control algorithm in a centralized unit. The use of a centralized LED with no intermediate electro-optical conversion stages for optical fiber-wireless links has been also recently proposed in factory networks.

BMW or Wieland Electric are examples of pioneer companies where LiFi technology has been successfully employed in production lines although further improvements are required before full deployment while ongoing joint research projects between companies and academia continue to explore the potential of VLC/LiFi systems in an industrial environment towards reliable, low latency and secure wireless communication for future Industry 4.0 scenarios.

### III. LOCALIZATION IN INDUSTRIAL ENVIRONMENTS

Smart manufacturing systems in Industry 4.0 will require precise location information to be able to know the positions of production resources (i.e., tools, robots, products, workers, etc.) and to deploy intelligent transport systems within the factory environments. For the outdoor environment, the global positioning system is widely used as a navigation system. However, its use inside buildings is limited by the severe attenuation suffered by RF signals from satellites. For indoor localization, there are localization techniques based on Wi-Fi, but the obtained accuracy is rather limited because of the complex multi-path propagation of radio waves. Ultra-wideband (UWB) radio is an alternative as these systems can obtain high accuracy, but high-cost implementation limits its application in manufacturing plants.

### IV. CHALLENGES AND FUTURE DIRECTIONS

OWC is a promising alternative to RF-based indoor localization, moreover, it can be combined with illumination when visible light is used for communications. The main advantages of this solution are (*i*) can be easily deployed in buildings; (*ii*) makes use of an unlicensed spectrum; (*iii*) multi-path has a negligible impact; and (*iv*) avoids interference into other rooms. There are two main alternatives to implementing OWC indoor localization: either based on optical camera communications or based on Li-Fi communications [2]. The first approach uses illumination fixtures to transmit a unique low rate (few kHz) beacon signal which can be captured by a camera to identify its localization, this information can be combined with the angle of arrival estimation. The Li-Fi alternative makes use of the standard G.9991 PHY frame structure to implement time-of-flight measurements to estimate the position of the receiver, it requires the receiver to have access to at least 3 transmitter units and gives higher accuracy than the first approach.

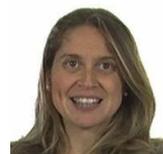

**Prof Beatriz Ortega** got her Ph.D. in Telecommunications Engineering in 1999 from the Universidad Politécnica de Valencia (UPVLC). She joined the Departamento de Comunicaciones at the Universidad Politécnica de Valencia in 1996, and she is a Full Professor since 2009. She has published more than 200 papers and conference contributions in fibre Bragg gratings, microwave photonics and optical networks and according to SCOPUS database, her H factor is 23. She has got several patents, and she is also a co-founder of EPHOOX company, which is a UPVLC spinoff. She has participated in more than 50 European and National Research Projects and Networks of Excellence. She is currently an Associate Editor of Journal of Lightwave Technology, and her current interests include millimetre wave signals, wireless optical systems and visible light communications.

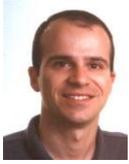

**Dr Vicenç Almenar** received the degree in Telecommunications Engineering and the Ph.D. degree from the Universitat Politècnica de València (UPV), in 1993 and 1999, respectively. In 2000, he did a Postdoctoral Research stay at the Centre for Communications Systems Research (CCSR), University of Surrey, U.K., where he was involved in research on digital signal processing for OFDM systems. He joined the Communications Department, UPV, in 1993, and now works as Full Professor. He was the Deputy Director of the department, from 2004 until 2014. His current research interests include OFDM, MIMO, and signal processing techniques for wireless and optical communications systems.

# 7. VIRTUAL REALITY

Olivier Bouchet
Orange Innovation, olivier.bouchet@orange.com

I. VIRTUAL REALITY STATE OF THE ART

Currently, there are two major players in the VR HMD market (Oculus, HTC) and two entrants (PiMax and Varjo). Oculus and HTC propose both tethered and untethered HMDs, untethered ones being able to embed computing (Oculus Go, Oculus Quest, HTC Vive Focus), or use wireless communication between the graphic server and the headset (HTC Vive Cosmos with a Vive wireless adapter). We have seen the development of Inside-out tracking solutions that do no longer need external to track the headset and its controllers. This inside-out tracking solution can theoretically make them mobile in a boundless space but they are unfortunately still tethered and rely on a computer for processing power.

A problem posed by tethered HMDs is the actual tether. Tethered HMDs require to be plugged into a computer in order to work. To address this problem, some companies like MSI or HP build VR backpacks that allow the players to carry the computer on their backs while running through the simulation. Wireless VR is even more important for user comfort considering not only the weight of the backpacks, but also the freedom of movement required by the convoluted custom stages built by LBE VR companies to maximize user immersion.
To address this problem, companies like TPCast provide external modules that can be plugged into the computer and the HMD to allow wireless video and audio transmission with extremely low latency [1]. The TPCast boasts latency under 2ms and uses a proprietary wireless protocol in the 60GHz band range for a data bandwidth of 7 Gbps. However, the problem with TPCast and other similar solutions is that they can only be used for one user at a time because it is impossible to get multiplex the signals on the 60GHz band.

There are currently three different classes of HMDs with varying screen resolutions: the *mass market* HMDs such as the Oculus Quest, the Oculus Rift S, the HTC Vive Cosmos, the HTC Vive Focus are currently at 1440x1600 (or 1440x1700) per eye; and *high-end mono panel* HMDs, such as the PiMax 8K with a resolution of 2560x1440 (with an upscaling on the screen with a perceived resolution of 3840x2160) pixels per eye; finally the *high-end multi-panels* HMDS such as the Varjo VR-2 where the perceived image is optically composed with two panels, one for the central field of view (1920x1080), and a second one for the peripheral field of view 1440x1600). HMD screen resolution is and will remain a major factor in the VR market competition and we believe the resolution will keep increasing with each generation until we reach the optimal resolution of 9000x7800 pixels per eye which is the theoretical perfect resolution.

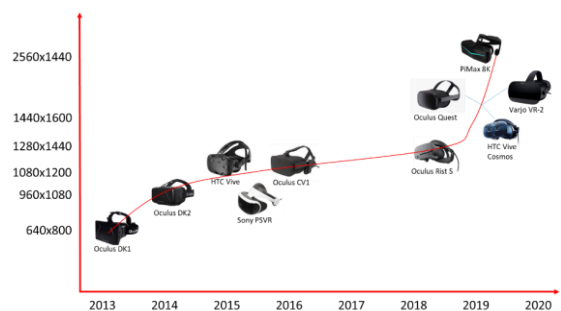

**Fig. 1.** Evolution of HMD resolution

Indeed, the growth of display resolutions raises the issue of real-time rendering, directly dependent on-screen resolution: the higher the display resolution, the longer the rendering takes. Besides increasing resolution, another barrier to fully immersive free-roaming VR is the restrained area in which a user can move. In the context of tethered HMDs, the largest area in which a user can currently be tracked is 4x4 meters using the HTC Vive lighthouse tracking system. It is possible to build custom solutions using the independent tracking systems developed for CAVE-like systems (such as ART or OptiTrack) but they must be custom built and custom software developed and externally integrated with existing systems.

II. VR AND OPTICAL WIRELESS SOLUTION

Virtual Reality is quickly becoming a fast growing market and the technology is fast evolving. In order to accommodate the growing market of multiuser VR, untethered VR is mandatory. But growing resolutions and graphics performance to enhance immersion of the end user means that more and more computer power will be needed to simulate realistic VR worlds in ultra-high resolutions of 9000x7800 pixels per eye in HDR. Furthermore, the low photon to motion latency required to avoid motion sickness implies that zero latency compression is needed to transmit the video data, resulting in the need for extremely high bandwidth needs.

At its most ideal resolution a bandwidth of 470 Gbps is needed per user in the case of uncompressed video, while zero latency encoding can bring it down to around 120Gbps par user. The key performance indicators (KPI) table below recaps the different bandwidth requirements for optimal wireless multiuser VR.

| Name | Requirement | | |
|---|---|---|---|
| | Mass Market VR | High End VR | Prototype VR |
| Potential HMD | HTC Vive Cosmos / Oculus Quest | Pimax 8K | N/A |
| Resolution | 2 × 1440 × 1700 | 2x3840×2160 (2×4K) | 2×7280×4320 (2×8K) |
| Bits per pixel | 24 | 30 (HDR) | 30 (HDR) |
| Frequency | 90 Hz | 120 Hz | 120 Hz |
| Connection density | 10 users / 100 m² | | |
| DL Data Rate per user | 9.8 Gbps | 55.62 Gbps | 210.88 Gbps |
| DL Traffic Density | 98.5 Gbps / 100 m² | 556.2 Gbps / 100 m² | 2.06 Tbps / 100 m² |
| Low latency video compression bandwidth for 10 users (4:1) | 24 Gbps | 139 Gbps | 527.2 Gbps |
| Video Stream Latency | < 2 ms | < 2 ms | < 2 ms |

**Table 1.** KPI and estimated requirements for wireless multiuser VR in 2021

This clearly shows the need for wireless Tbps or Tbit/s connections and beyond in order to allow the spread and growth of multiuser VR applications. To achieve this target and in the context of WORTECS (Wireless Optical / Radio TErabit CommunicationS) European collaborative project [2], Orange Labs demonstrated in the University of Southampton's Optoelectronics Research Centre (ORC) and in partnership with Oxford University, the possibility of wireless optical communication at 1.16 Tbit/s [3]. Using an optical medium of ten 100 Gbit/s wavelengths, the demonstration was able to achieve an aggregated rate of 1 Tbit/s (1000 Gbit/s) in both directions in an enclosed space. Irrespective of the world record, this wireless system offers photonic data speeds of 1 Gbit/s to 1 Tbit/s.

To proceed in a room, take a light beam from an Access Point (AP) optical fiber and automatically direct it towards the User Terminal (UT), then proceed reciprocally in order to obtain ultra-high speed communication in both directions. These Fi-Wi terminals include a light concentrator and a location and guidance unit for the beam so that it can follow the user's movement. This enables a direct connection between two optical fibers, with no need to transform the optical signal into an electrical signal (or vice versa), and no need for the signal processing and regeneration required for the radio alternative. Fi-Wi connections are intrinsically bidirectional, transparent to modulation type, to protocol and to wavelength, with a speed of 1 Gbit/s to 1 Tbit/s (theoretical maximum currently assessed at 44.2 Tbit/s), without modification to access point or user terminal. Data transmission is highly secured with a very closely directed light beam. Electricity consumption is low and connectivity benefits from guaranteed speeds for each user since this is point-to-point communication.

So Fi-Wi is a technological breakthrough, linking the fiber optics network directly to terminals. The solution creates an opportunity for an end-to-end optical network that uses fiber and remains wireless. Research in the field is generating increasing interest, and one of the next steps will be to go further with innovative applications, particularly in multi-user mode.

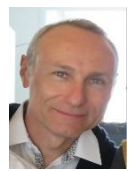

M. Olivier Bouchet (m) received the Licence Es Physical Science and the Telecommunication Engineer Diploma in 1987 from University of Rennes and 1989 from Polytech of Angers respectively. A Master of Business of Administration (MBA), in 1992 from University of Rennes, completed his studies. His first activity, in France Telecom, was project leader for radio paging mass market product. He joined Orange Labs in 1998. His current research interests are in the field of Optical Wireless Communications (OWC), Light Fidelity (LiFi), Free Space Optic (FSO) and Fiber Wireless (FiWi). He is author or coauthor of 5 books, around 60 papers or oral communications and holds 34 patents. He was European OMEGA Work Package leader on Optical

Wireless communication, then ECONHOME French project and ACEMIND European Celtic Plus project leader. He is currently European Community H2020 WORTECS project leader.

# 8. IPV6-BASED IOT IN 2025

Latif Ladid
Secan-Lab, University of Luxembourg, Luxembourg, latif.ladid@uni.lu

## I. INTRODUCTION

The number of Internet Connected devices will cross the incredible total of 50 billion by 2025. The connectivity fabric of IP is used to enable more and more efficient context exchange with a broader range of devices and things. This results in the Internet of Things (IoT). Projected to increase device counts by orders of magnitude over the next few decades, IoT's impact cannot be overstated. Already enabling a rich set of new capabilities in Smart Cities, Smart Grid, Smart Buildings, and Smart Manufacturing, IoT stands to transform virtually every part of modern life that automation or visibility may improve.

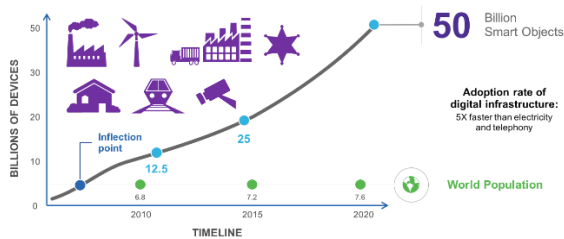

Source Cisco

**Fig. 1**. IoT growth

## II. IOT CONNECTIVITY: WIRED AND WIRELESS

No matter the precise forecast, the sheer tsunami of devices coming online in the next months, years, and decades ensures that the future is not exclusively, or even significantly, wired. Wireless with its adaptability and ease will inevitably dominate the IoT landscape. Exactly which wireless technology or technologies will be used remains relatively unclear, as many new technologies are still emerging, while others are still early in the standards process.

The challenges IPv6 poses to high bandwidth wireless networks are well-known. However, low bandwidth links, like LPWAN (Low Power Wide Area Network), do require optimization and broadly adapt and adopt techniques like IPv6 header compression. Clause 4.4 is describing the IETF technologies to adapt IPv6 to different constraint media. This problem is not specific to the use of IPv6 but is due primarily to the scale of IoT deployment.

The following list summarizes the main different wireless technologies used for IoT:

- IEEE 802.15.4 [1] WPAN: The IEEE 802.15 TG4 was chartered to investigate a low data rate solution with multi-month to multi-year battery life and very low complexity. It is operating in an unlicensed, international frequency band. Potential applications are sensors, interactive toys, smart badges, remote controls, and home automation.
- IEEE 802.11 [1]
- WLAN (Wireless Local Area Network).
- LPWAN (Low Power and Wide Area Network).
- Cellular Networks (NB-IoT, 5G).

## III. MOTIVATION FOR IPV6 IN THE IOT

i) Main driver

The main driver is probably the large address space that IPv6 is providing but it is not the only aspect: Auto-configuration, security, and flow identification bring huge advantages to IoT systems as well as being a future-proof technology.

ii) Addressability

Global, public, and private address space have been defined for IPv6; therefore, a decision must be made regarding which type of IPv6 addressing scheme should be used. Global addressing means you should follow the Regional Internet Registries (RIR) policies (such as ARIN https://www.arin.net/policy/nrpm.html) to register an IPv6 prefix that is large enough for the expected deployment and its expansion over the coming years. This does not mean the address space allocated to the infrastructure has to be advertised over the Internet allowing any Internet users to reach a given device.

The public prefix can be advertised if representing the entire corporation - or not - and proper filtering mechanisms are in place to block all access to the devices. On the other end, using a private address space means the prefix not be advertised over the Internet, but, in case there is a need for Business-to-Business (B2B) services and connectivity, a private address would lead to the deployment of additional networking devices known as IPv6-IPv6 NPT (Network Prefix Translation, IETF RFC 6296 [**Error! Reference source not found.**]) gateways.

Three methods to set an IPv6 address on an endpoint are available:

- **Manual configuration:** This method is appropriate for headend and NMS servers that never change their address, but is inappropriate for millions of end-points, such as meters, because of the associated operational cost and complexity.

- **Stateless auto configuration:** This mechanism is similar to Appletalk, IPX, and OSI, meaning an IPv6 prefix gets configured on a router interface (interface of any routing device such as a meter in a mesh or PLC AMI network), which is then advertised to nodes attached to the interface. When receiving the prefix at boot time, the node can automatically set up its IPv6 address.

- **Stateful auto configuration:** Through the use of Dynamic Host Control Protocol for IPv6 (DHCPv6) Individual Address Assignment, this method requires DHCPv6 server and relay to be configured in the network. It benefits from strong security because the DHCPv6 process can be coupled with authentication, authorization, and accounting (AAA), plus the population of Domain Name System (DNS) available for headend and NMS applications.

iii) Security Mechanism

In the past, it was sometimes claimed that the use of open standards and protocols may itself represent a security issue, but this is overcome by the largest possible community effort, knowledge database, and solutions available for monitoring, analysing, and fixing flaws and threats - something a proprietary system could never achieve.

iv) IP up to the end device/end to end principle

The past two decades, with the transition of protocols such as Systems Network Architecture (SNA), Appletalk, DECnet, Internetwork Packet Exchange (IPX), and X.25, showed us that such gateways were viable options only during transition periods with smaller, single-application networks. But proprietary protocol and translation gateways suffer from well-known severe issues, such as high capital expenditures (CapEx) and operating expenses (OpEx), along with significant technical limitations, including lack of end-to-end capabilities in terms of QoS, fast recovery consistency, single points of failure (unless implementing complex stateful failover mechanisms), limiting factors in terms of innovation (forcing to least common denominator), lack of scalability, vulnerability to security attacks, and more. Therefore, using IPv6 end to end (that is, IP running on each and every device in the network) will be, in many ways, a much superior approach for multiservice IoT networks. See IETF RFC 3027 [2] as an example of protocol complications with translation gateways.

IV. CONCLUSIONS

IPv6 can enable and sustain the growth rate of the IoT. It offers a future-proof solution. More and more SDOs (Standardization Development Organization) have decided to either transition to IPv6 or to develop new standards only based on IPv6. This is specifically the case for IoT-related standards. 3GPP secretary and CTO of ETSI Adrian Scrase has already announced back in April 2019 the move from E.164 for Machine Type communication to IPv6 addressing for larger-scale deployment of IoT. IPv6 does not only enable the scalability required by the IoT but also provides enhancement from IPv4 in the field of mobility support, stateless address auto-configuration, support of constraint devices, and security to mention only a few of them.

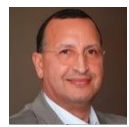
**Latif Ladid** is Founder & President, IPv6 FORUM and Chair, ETSI IPv6 Industry Specification Group. He is the Founding co-chair, IEEE 5G World Forum and Co-Chair, IEEE GET Blockchain Forum. He is active in the IoT Forum as Board Member and Chair of the Global IoT Summit and * Member of 3GPP PCG (Board). Latif held previous voluntary positions as Former Chair, IEEE IoT World Forum, Former Chair, IEEE ComSoC IoT subcommittee, Former Chair, IEEE ComSoC 5G subcommittee, Former Vice Chair, IEEE ComSoC SDN-NFV subcommittee, Emeritus Trustee, Internet Society - ISOC and Emeritus World summit Award Board Member. Latif is currently Research Fellow @ University of Luxembourg on multiple European Commission Next Generation Technologies Projects and Member of Future Internet Forum EU Member States (representing Luxembourg)